\documentclass[letterpaper, 10 pt, conference]{ieeeconf}  

\IEEEoverridecommandlockouts                              

\usepackage{cite}

\usepackage[pdftex]{graphicx}
\graphicspath{{Figures/}{Figures/ExampleImages/}{Figures/CNNs/}}
\DeclareGraphicsExtensions{.pdf,.jpeg}
\usepackage{amsmath, amsfonts}
\usepackage{algorithmic}

\usepackage[caption=false,font=footnotesize]{subfig}

\usepackage{url}

\newcommand{\includeCroppedPdf}[2][]{%
    \IfFileExists{./#2-crop.pdf}{}{%
        \immediate\write18{pdfcrop #2 #2-crop.pdf}}%
    \includegraphics[#1]{#2-crop.pdf}}

\title{\LARGE \bf UFRC: A Unified Framework for Reliable COVID-19 Detection on Crowdsourced Cough Audio }

\author{Jiangeng Chang$^{1,2,\ast}$, Yucheng Ruan$^{1,2,\ast}$, Cui Shaoze$^{2}$, John Soong Tshon Yit$^{4,5,\dagger}$, Mengling Feng$^{1,2,3,\dagger}$
\thanks{$^{1}$Saw Swee Hock School of Public Health, National University of Singapore}%
\thanks{$^{2}$National University Health System, National University of Singapore}%
\thanks{$^{3}$Institute of Data Science, National University of Singapore}%
\thanks{$^{4}$Department of Medicine, National University Hospital, Singapore}%
\thanks{$^{5}$Yong Loo Lin Medical School, National University of Singapore, Singapore}%
\thanks{$^{\dagger}$Send correspondence to \{john\_soong, ephfm\}@nus.edu.sg}%
\thanks{$^\ast$  Contributed equally}%
}

\begin{document}

\maketitle
\thispagestyle{empty}
\pagestyle{empty}

\begin{abstract}
We suggested a unified system with core components of data augmentation, ImageNet-pretrained ResNet-50, cost-sensitive loss, deep ensemble learning, and uncertainty estimation to quickly and consistently detect COVID-19 using acoustic evidence. To increase the model's capacity to identify a minority class, data augmentation and cost-sensitive loss are incorporated (infected samples). In the COVID-19 detection challenge, ImageNet-pretrained ResNet-50 has been found to be effective. The unified framework also integrates deep ensemble learning and uncertainty estimation to integrate predictions from various base classifiers for generalisation and reliability. We ran a series of tests using the DiCOVA2021 challenge dataset to assess the efficacy of our proposed method, and the results show that our method has an AUC-ROC of 85.43 percent, making it a promising method for COVID-19 detection. The unified framework also demonstrates that audio may be used to quickly diagnose different respiratory disorders.
\end{abstract}




\section{Introduction}
By September 22nd 2021, the total number of Coronavirus disease 2019 (COVID-19) confirmed cases has reached 230 million worldwide, and unfortunately, the pandemic is still ongoing. The Reverse Transcription Polymerase Chain Reaction (RT-PCR) is the current gold standard for COVID-19 screening.
However, the RT-PCR is costly in terms of time, manpower and resources \cite{cevik2020virology, vogels2020analytical}. For remote and less developed regions, it can be difficult for people to afford large scale TR-PCR tests to detect all cases. Even for developed regions, the costly nature of RT-PCR tests can also lead to considerable delays in diagnosis when facing a large number of suspicious cases.
Therefore, researchers are constantly searching for more cost-effective and easy-to-access test methods that ideally can identify the infected individuals on the spot. It then comes to researchers' attention that cough is one of the most common respiratory symptoms in the early stage of infections. Some study has also shown that cough audios can be potentially adapted for quick diagnoses of COVID-19~\cite{laguarta2020covid,coppock2021covid,xia2021uncertainty}.

Since respiratory audios are relatively easy to obtain at a lower cost, we envision that the respiratory audio-based diagnosis approach can be a fast and cheap COVID-19 detection solution for rural and underdeveloped areas \cite{vogels2020analytical}.
At present, researchers have developed some machine learning\cite{pahar2021covid} or deep learning \cite{laguarta2020covid,pal2021pay,casanova2021deep} algorithms to diagnose COVID-19 through the respiratory audio. Their success, to a certain degree, has proved the possibility of detecting COVID-19 through audio. 

Although the great performance of algorithms demonstrates that audio-based detection of COVID-19 can be considered an effective method, the datasets used to develop these algorithms are collected by individual research groups or institutions, which might not be easily accessed by other researchers to investigate and validate new methodologies.

Unlike clinical datasets above, crowdsourced data is a new type of data that uses the existing web environment to collect data from volunteers in different regions, allowing a large amount of experimental data to be obtained at the beginning of a respiratory epidemic such as COVID-19. These data can be used as a form of open-source data to facilitate a wide range of research for a variety of tasks.

However, these data also have several serious drawbacks. The first one is the data imbalance. Since crowdsourced data is based on voluntary contributors, it is difficult to keep a balance between positive and negative samples. To achieve balance, randomly removing negative samples or replicating positive samples makes it difficult to generate a robust model. In addition, crowdsourced data is collected on online platforms, leading to the discrepancy in data quality in different circumstances.

\begin{figure}[h]
  \centering
  \includegraphics[width=\linewidth]{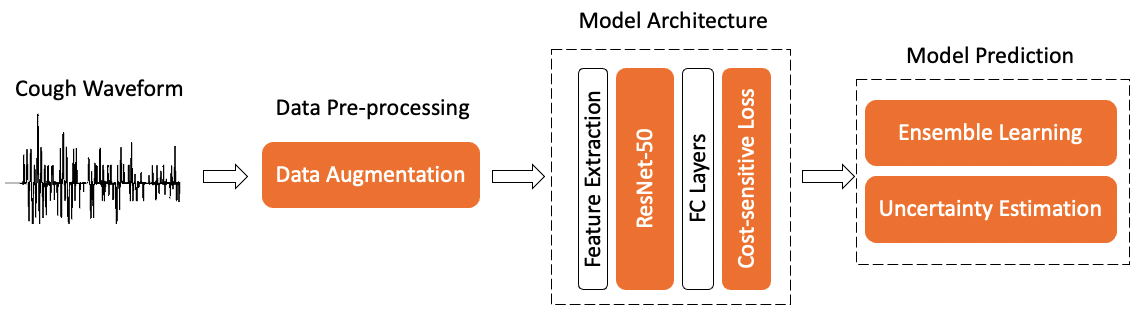}
  \caption{The illustration of our proposed frameworks and the key components have been colored with orange.}
  \label{fig:framework_1}
\end{figure}

In this study, we proposed a unified framework for rapidly diagnosing COVID-19 using crowdsourced data.  The key components of the framework (Figure \ref{fig:framework_1}) include: 
(A) \textit{Data Augmentation} Data augmentation had been used in our framework by adding some random noises to the audio to produce new data in the same label and it can help in the circumstances of datasets with a limited number of positive samples. 
(B) \textit{ImageNet-pretrained ResNet-50} Even though there are significant differences between the Mel spectrograms and images, a lot of research work has explored the application of ImageNet-Pretrained ResNet-50 to the field of audio speech for finetuning \cite{gong2021psla,palanisamy2020rethinking}. We also here experimentally investigated that pre-trained ResNet-50 weights lead to a significant improvement in crowdsourced COVID-19 detection framework compared to randomly initialised parameters.
(C) \textit{Cost-sensitive Loss} Cost-sensitive learning focuses more on the costs of prediction errors from minor classes when training a machine learning model. 
(D) \textit{Deep Ensemble Learning} Deep ensemble learning method has been deployed to integrate different base classifiers for better model generalizability.  
(E) \textit{Uncertainty Estimation} The qualified uncertainty can be used for selective prediction: keeping low-uncertain outputs but referring high uncertain (unsafe) predictions to authoritative doctors for external checks, which can help to improve the system robustness.   

\section{Methods}
In this section, we will illustrate the key components in our general framework for the COVID-19 detection task.

\subsection{Gussian Noise-based Data Augmentation}


To increase the diversity of the training datasets, data augmentation is frequently required. Furthermore, data augmentation can reduce the domain mismatch between the enrolled and test data, according to \cite{Jiang2020TheXS}. Flip, rotation, scale, crop, translation, and other data augmentation techniques are common. For data augmentation, we chose to add Gaussian noise to the raw data in our investigation.

A pseudorandom number generator can generate Gaussian noise, which has a mean of zero and a standard deviation of one. Gaussian noise was used to produce additional synthetic minority samples for our proposed model, which helps to lower the incidence of overfitting in DNNs.

\subsection{Finetuning on ImageNet-Pretrained ResNet-50}
ResNet has proven to be a powerful backbone in the field of audio classification. However, due to the model's large number of parameters, learning on short datasets with random initial parameters is insufficient. As a result, much audio classification research has attempted to finetune parameters from ImageNet-pretrained ResNet-50, and this method has been shown to deliver significant improvements in audio tagging\cite{gong2021psla}, audio classification\cite{palanisamy2020rethinking}, and environmental audio classification\cite{9533654}.

On the DiCOVA dataset, we fine-tuned the ImageNet-pretrained ResNet-50-based backbone inspired by the previous work.

\subsection{Cost-sensitive Loss (Focal Loss)}
In the training of deep learning models, the loss function is used to measure the degree of difference between the predicted values and the ground truth values. The loss function plays the role of "supervisor" in DNNs, which guides the model training to the global minimum. Typically, Cross Entropy (CE) is often used as the loss function, which is calculated as follows Eq.(\ref{eq5}). 

\begin{equation}
  L_{CE} = -\sum_{i=1}^{m}{y_i\cdot{log(p_i)}}
  \label{eq5}
\end{equation}

where $y_i$ represents the label of sample $i$, and $p_i$ is the probability that sample $i$ is predicted to be positive.

Cross Entropy can help to generate a good model when the number of samples doesn't differ so much in each class, but it is no longer effective when the data is imbalanced.

For this, Lin et al. \cite{lin2017focal} proposed a new loss function called Focal Loss which is calculated as Eq.(\ref{eq6}).

\begin{equation}
  L_{FL} = -\sum_{i=1}^{m}{\alpha_i(1-p_i)^\gamma {log(p_i)}}
  \label{eq6}
\end{equation}

where $\gamma$  $(\gamma \geq 0)$ is the focusing parameter which is used to adjust the weights of difficult samples and easy samples, and the $(1-p_i )^\gamma$ is called the modulating factor. In addition, $\alpha_i$ is used to adjust the weights of positive and negative samples.

From Eq.(\ref{eq6}), there are two important properties of the Focal Loss: (1) When the value of $p_i\rightarrow1$, $(1-p_i )^\gamma \rightarrow0$, it indicates that the prediction of the model is accurate and the contribution of these easy samples to the loss is quite small. (2) The focusing parameter $\gamma$ smoothly adjusts the rate at which easy examples are down-weighted. 

When $\gamma=0$, Focal Loss is equivalent to CE, and as $\gamma$ increases, the weights of easy samples will be further reduced. Therefore, using Eq.(\ref{eq6}) as the loss function of the model in this paper, the problem of data imbalance will be greatly alleviated.

\subsection{Deep Ensemble Learning} To enhance the predictive performance of independent trained identical models, we introduce an ensemble model to overcome the problem of diversity introduced by differences in initialization and mini-batch orderings.  


It appears that a random initialization of the NN parameters as well as random shuffling of the data points is sufficient for obtaining good performance \cite{lakshminarayanan2016simple,chang2021dicova}.

\begin{figure}[h]
  \centering
  \includegraphics[width=0.9\linewidth]{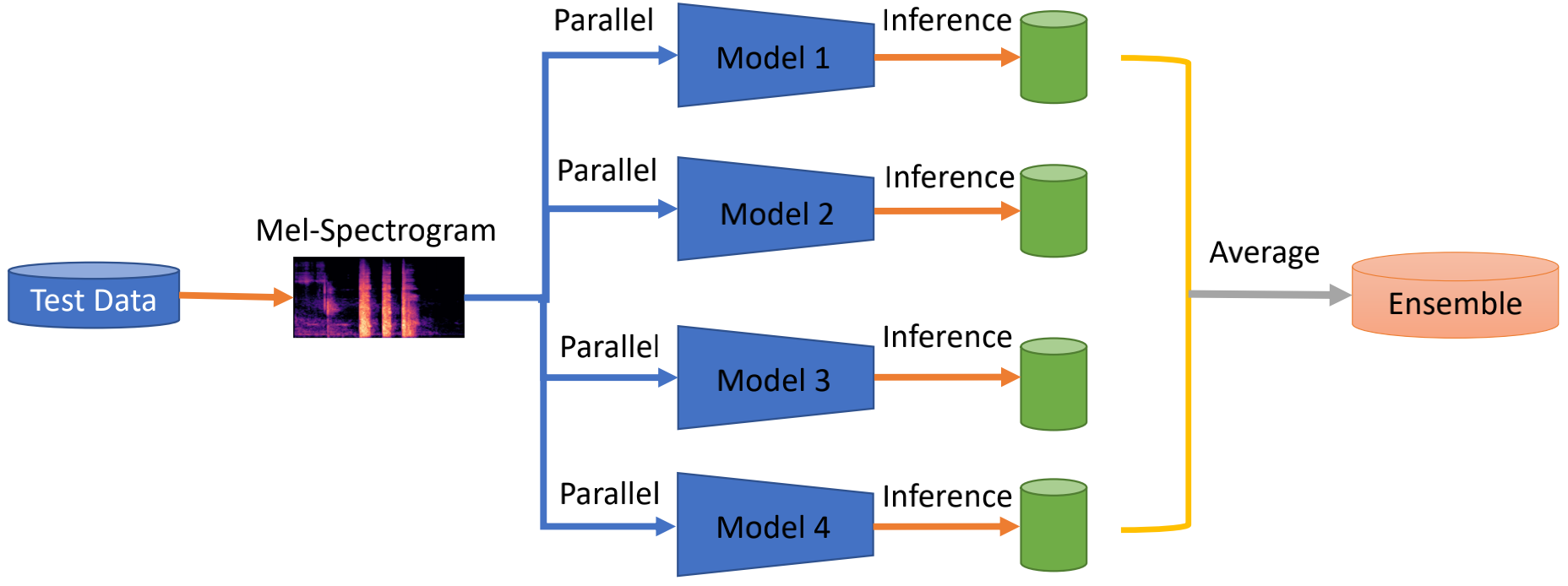}
  \caption{The models used in deep ensemble all have the same architecture, with the variability coming from the order in which the training data is passed during training, so that multiple models can be predicted in parallel during prediction and the results are averaged and integrated before prediction.}
  \label{fig:framework_2}
\end{figure}


Figure \ref{fig:framework_2} shows our ensemble learning techniques, and we used the average bagging method in ensemble learning to integrate base learners for model calibration. In order to predict audio classification scores, we trained $M$ independent models using the same architecture, hyperparameter settings, and training procedures. To calculate the final probability, Eq.(\ref{eq7}) is calculated as the average of soft-max outputs of these $M$ individually trained models.
\begin{equation}
  y_{final} = \frac{1}{M}\sum_{m=1}^{M}{y_m} 
  \label{eq7}
\end{equation}

where $y_m$ is the soft-max output from individual model.

\subsection{Reliable Estimation}
Following the research \cite{xia2021uncertainty}, we also identified the level of disagreement across models within the ensemble suite as the measure of uncertainty because softmax probability was unable to convey model confidence. Deep ensemble uncertainty has also been found to be more effective than other estimating methodologies~\cite{lakshminarayanan2016simple}. 
If the uncertainty is greater than a predetermined threshold, the model's prediction during digital pre-screening is considered reliable.
\section{Experiments and Results}
In this section, we aim to address the following questions:

\noindent$\bullet$ \textbf{Q1:} Do data balancing techniques and deep ensemble learning contribute to model performance improvement?

\noindent$\bullet$ \textbf{Q2:} Is it ImagNet-pretrained ResNet-50 better?

\noindent$\bullet$ \textbf{Q3:} Is uncertainty of proposed models reliable?

\noindent\textbf{Dataset} To be objective, we evaluated our model on the DiCOVA2021 Task 1 dataset\footnote{\url{https://dicova2021.github.io/#home}}. The challenge organizers have already split the dataset into train set and validation set. The train set contains 822 samples which includes 772 non-COVID-19 ones and 50 COVID-19 ones. The validation set contains 218 samples which includes 193 non-COVID-19 ones and 25 COVID-19 ones.\\
\noindent\textbf{Evaluation Metrics} The area under the receiver operating characteristics curve, or \textit{ROC-AUC}, was generated to further analyze the effectiveness of our approach. In addition, the mean and standard deviation of the aforementioned parameters were provided across 5-fold runs.

\noindent\textbf{Hyperparameter}. For training, we set the number of iterations with \textit{20} epochs and used \textit{Adam} as the optimizer. The batch size is \textit{16}.

\subsection{Results and Discussions}

Three sets of experiments were conducted to evaluate the performance of the proposed model. Firstly, to demonstrate the usefulness of the data balancing technique in our model, we first exhibit the results of the various components of the proposed model on the test set. Secondly, we evaluated the differences in model performance between random initial weights and ImageNet-pretrained weights. Lastly, we investigated how the deep ensemble methods help to calibrate the prediction results. 

\subsubsection{Data Balancing Techniques and Deep Ensemble Learning for Model Performance}
We compared the performance of the ImageNet-pretrained ResNet-50-based model using two different loss functions (Cross Entropy (CE) and Focus Loss (FL)) and two different data augmentation methods (Simple Duplicated Minor Class (Dul) and Gaussian Noise (Gua)).

In addition, we also added three baselines given by DiCOVA2021 into comparison, all of which are based on three traditional classifiers, SVM, Random Forest and Multilayer Perceptron using traditional acoustic MFCC features.

\begin{figure}[h]
  \centering
  \includegraphics[width=0.9\linewidth]{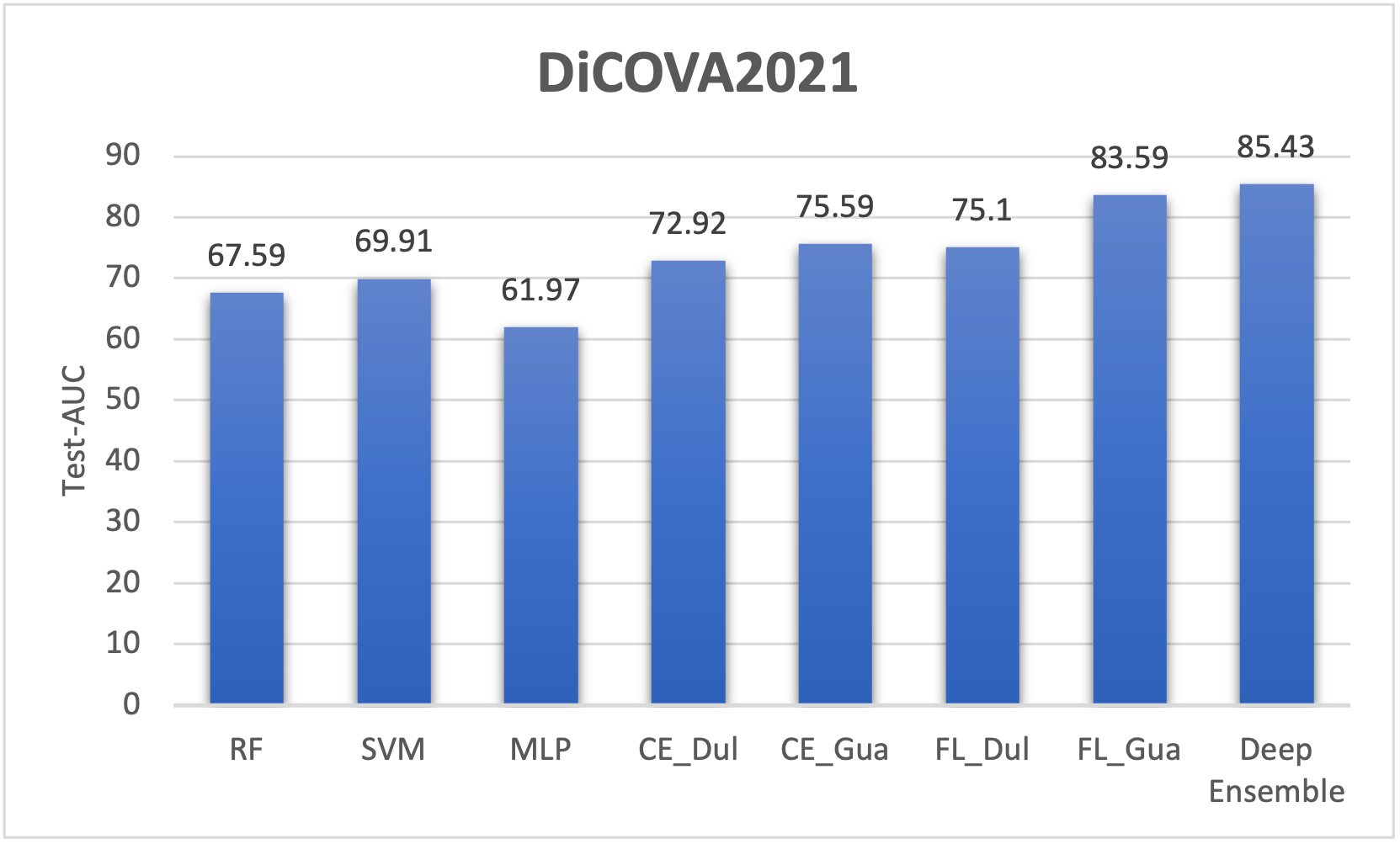}
  \caption{DiCOVA2021 Test Result}
  \label{fig:framework_3}
\end{figure}

As is shown in Figure~\ref{fig:framework_3}, the performance of all types of our proposed ImageNet-pretrained ResNet-50-based models are better than the challenge’s official baselines. In terms of the loss function, the model with CE performs better than that with FL on the test set, and overfitting occurs when FL is used for training on the original training set. However, after using Dul or Gua for data augmentation, the model with FL performed better than that with CE. It proves that a combination of multiple data balancing techniques will contribute more to the model performance. Moreover, it can be observed that using Gua to process imbalanced data is better than directly replicating minority samples.

Deep ensemble learning has also been applied with FL and Gua to generate the calibrated model. The results in Figure \ref{fig:framework_3} shows the the contribution of deep ensemble approach to the performance improvement.

\begin{table}[h]
\renewcommand{\arraystretch}{1.3}
\caption{Results on DiCOVA2021 Validation dataset}
\label{tab:trainingSummary}
\centering
\begin{tabular}{cc}
\hline
Methods & Test AUC \\
\hline 
Random\_initial &  43.48\% \\
Imagenet\_initial &  64.50\% \\
 \hline 
\end{tabular}
\end{table}

\subsubsection{Random Initial Weights vs ImageNet-pretrained Weights}
\begin{figure}[h]
  \centering
  \includegraphics[scale=0.5]{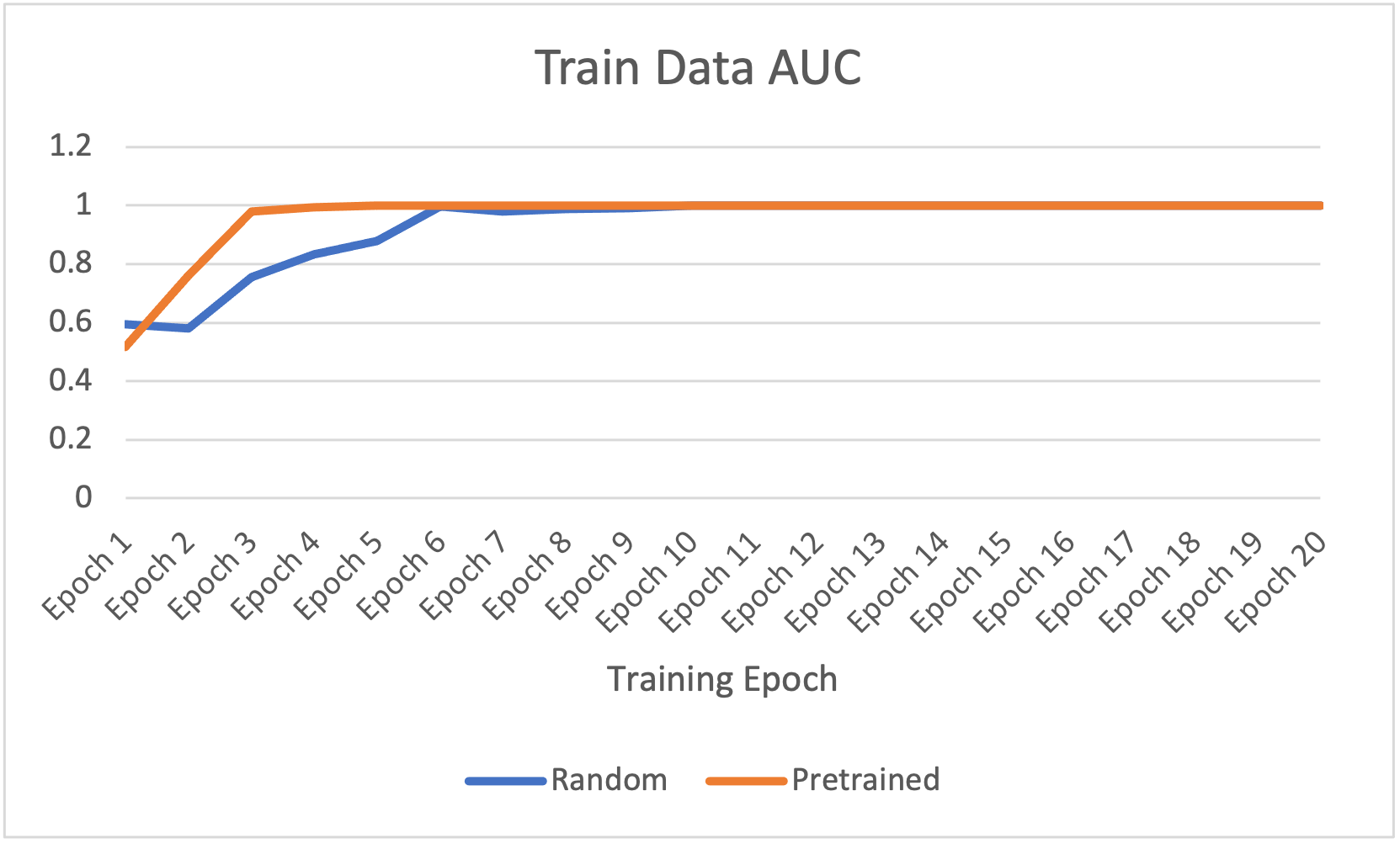}
  \caption{Train Dataset AUC changes in Every Epoch}
  \label{fig:framework_4}
\end{figure}

In this section, we further experimentally compared the performance of the  ResNet-50-based model with random initial weights and the ResNet-50-based model with ImageNet-pretrained initial weights on the dataset, where we randomly selected 20\% of the entire dataset (train+val) as the test set, 15\% of the dataset as the validation dataset, and the rest as the training set. The performance on the test dataset shows that the ResNet-50-based model with ImageNet-pretrained initial weights can substantially improve the performance of the model. Additionally, the curves of the models with two different weight initialization (Figure \ref{fig:framework_4}) show that although the small training dataset allows both initialization methods to achieve an AUC of 1 in less than 10 epochs, the ResNet-50-based model with ImageNet-pretrained initial weights converges faster. Thus, it can be concluded that using ImageNet-pretrained initialization has a huge advantage in generalization and converging time over random initialization of weights.

\subsubsection{Estimation and Application of Uncertainty}
In this section, we estimated the uncertainty of the prediction results of our proposed framework. According to Table~\ref{tab:reliable}, we divided the test dataset into two comparison datasets: the one with low uncertainty and the other one with high uncertainty, according to the distribution of uncertainty of the proposed model's prediction results in the test set. According to the accuracy of test set prediciton, we found that the accuracy in the test dataset with low uncertainty were 0.91 compared to 0.86 for that with high uncertainty. In practice, the predictions of the model can be selected based on the uncertainty from our proposed model to perform reliable COVID-19 diagnoses.
\begin{table}[h]
\renewcommand{\arraystretch}{1.3}
\caption{Uncertainty Comparison Result}
\label{tab:reliable}
\centering
\begin{tabular}{ccc}
\hline
 & Low Uncertainty & High Uncertainty \\
\hline 
Case Number & 119 & 99 \\
Accuracy &  0.91 & 0.86 \\
 \hline 
\end{tabular}
\end{table}

\indent

\section{Conclusion}

In order to identify patients infected with COVID-19 more accurately using audio data, we proposed a unified framework for reliable COVID-19 detection that incorporates multiple useful technologies. First, Gaussian noise-based data augmentation and Focal Loss were introduced to deal with imbalanced data. Based on the ResNet-50 pre-trained on ImageNet, we integrated the fine-tuning techniques with transfer learning to adjust the weights of the deep neural network for COVID-19 detection. In addition, in order to make our proposed model more robust and generalizable, we adopted ensemble learning and uncertainty estimation to integrate the predictions from multiple base models. Our experimental results show that the proposed method can effectively identify the infected persons with COVID-19 and is superior to other state-of-the-art methods.

The fast diagnoses of other respiratory diseases might also benefit from this unified framework and we leave this challenging extension for future work.

\section*{Acknowledgment}
We are very grateful to the organizers of the DiCOVA 2021 Challenge for their efforts in providing the participants with data and a platform for the competition. This research is partially supported by the National Research Foundation Singapore under its AI Singapore Programme (Award Number: [AISG-100E-2020-055 and AISG-GC-2019-001-2A]).

\bibliographystyle{IEEEtran}
\bibliography{refs}

\end{document}